\documentclass[twoside,twocolumn]{article}

\usepackage{tabulary,graphicx,times,caption,fancyhdr,amsfonts,amssymb,amsbsy,latexsym,amsmath}
\usepackage[utf8]{inputenc}
\usepackage{url,multirow,morefloats,floatflt,cancel,tfrupee,textcomp,colortbl,xcolor,pifont}
\usepackage[nointegrals]{wasysym}
\usepackage[left,switch,pagewise,displaymath,mathlines]{lineno}

\makeatletter

\usepackage{ifxetex}
\ifxetex\else
  \usepackage{dblfloatfix}
\fi

\@ifundefined{subparagraph}{
\def\subparagraph{\@startsection{paragraph}{5}{2\parindent}{0ex plus 0.1ex minus 0.1ex}%
{0ex}{\normalfont\small\itshape}}%
}{}

\def\URL#1#2{\@ifundefined{href}{#2}{\href{#1}{#2}}}

\def\UrlOrds{\do\*\do\-\do\~\do\'\do\"\do\-}%
\g@addto@macro{\UrlBreaks}{\UrlOrds}

\makeatother

\usepackage[paperheight=11in,paperwidth=8.3in,margin=2.5cm,headsep=.7cm,top=2.5cm]{geometry}
\usepackage[T1]{fontenc}

\renewenvironment{abstract}
	{\trivlist\item[]\leftskip0pt\par\vskip4pt\noindent
  	\textbf{\abstractname}\mbox{\null}\\}
	{\par\noindent\endtrivlist}

\def\keywords#1{\par\medskip\par\noindent\textbf{Keywords}: #1\par}

 \date{} \emergencystretch 8pt

\makeatletter
\def\author#1{\gdef\@author{\hskip-\tabcolsep%
	\parbox{\textwidth}{\raggedright\bfseries#1\\[1pc]}}}
\def\address[#1]#2{\g@addto@macro\@author{\\\hskip-\tabcolsep\parbox{\textwidth}{\raggedright%
	\normalsize\normalfont\textsuperscript{#1}#2}}}
\let\addresslink\textsuperscript
\def\correspondence#1{\g@addto@macro\@author{\\\hskip-\tabcolsep\parbox{\textwidth}{\raggedright%
	\vspace*{10pt}\normalsize\normalfont~\\#1~\\[12pt]}}}
\def\email#1{\g@addto@macro\@author{\\\hskip-\tabcolsep\parbox{\textwidth}{\raggedright%
	\normalsize\normalfont Emails: #1}}}

\def\title#1{\gdef\@title{\vspace*{-30pt}%
	\raggedright\textbf{\@journaltitle}~\\%
  \raggedright\bfseries\ifx\@articleType\@empty\vspace*{20pt}\else%
  \vspace*{20pt}\@articleType\vspace*{20pt}\\\fi#1}}
\let\@journaltitle\@empty \def\journaltitle#1{\gdef\@journaltitle{{\normalfont\itshape#1}}}
\let\@articleType\@empty \def\articletype#1{\gdef\@articleType{{\normalfont\itshape#1}}}

\let\@runningHead\@empty \def\RunningHead#1{\gdef\@runningHead{{\normalfont #1}}}

\usepackage{fancyhdr}
\fancypagestyle{headings}{\fancyhf{}
  \fancyhead[R]{\itshape\@runningHead}
  \fancyfoot[C]{\thepage}}
\fancypagestyle{plain}{%
	\fancyhf{}\fancyhead[R]{Wiley | Hindawi}
  \fancyfoot[C]{\thepage}}
\makeatother

\usepackage[%
	numbers,sort&compress%
  ]{natbib}

\usepackage{float,xcolor}

\usepackage{amsmath,amsfonts}
\usepackage{textcomp}
\usepackage{url}
\usepackage{verbatim}
\usepackage{graphicx}
\usepackage{booktabs}
\usepackage{tikz}
\usetikzlibrary{shapes.geometric,positioning}
\usepackage{color}
\usepackage{eurosym}
\newtheorem{theorem}{Theorem}
\def\UrlBreaks{\do\/\do-\do_}

\usepackage{nomencl}
\makenomenclature

\journaltitle{International Journal of Energy Research}

\begin{document}

% Title of the document
\title{Aggregated demand flexibility prediction of residential thermostatically controlled loads and participation in electricity balance markets}

% Author names
\author{%
		Alejandro Mart\'in-Crespo\addresslink{1},
  	Enrique Baeyens\addresslink{2},
  	Sergio Saludes-Rodil\addresslink{1} and
		Fernando Frechoso-Escudero\addresslink{3}
    }
		
% Affiliation
\address[1]{Centro Tecnol\'ogico CARTIF, Parque Tecnol\'ogico de Boecillo 205, Boecillo, 47151, Spain}
\address[2]{Instituto de las Tecnolog\'ias Avanzadas de la Producci\'on, Universidad de Valladolid, Paseo Prado de la Magdalena s/n, Valladolid, 47011, Spain}
\address[3]{Departamento de Ingenier\'ia El\'ectrica, Universidad de Valladolid, Paseo Prado de la Magdalena s/n, Valladolid, 47011, Spain}

% Corresponding author details
\correspondence{Correspondence should be addressed to 
    	Alejandro Mart\'in-Crespo: alemar@cartif.es}
			
% Emails of authors
\email{enrique.baeyens@uva.es (Enrique Baeyens), sersal@cartif.es (Sergio Saludes-Rodil), frechoso@eii.uva.es (Fernando Frechoso-Escudero)}%

% Running Head
\RunningHead{International Journal of Energy Research}

\maketitle 

% Abstract
\begin{abstract}
The aggregate demand flexibility of a set of thermostatically controlled
residential loads (TCLs) can be represented by a virtual battery (VB) in order
to manage their participation in the electricity markets. For this purpose, it
is necessary to know in advance and with a high level of reliability the
maximum power that can be supplied by the aggregation of TCLs. A probability
function of the power that can be supplied by a VB is introduced in this paper. 
This probability function is used to predict the demand flexibility using a 
rigorous experimental probabilistic method based on a combination of Monte Carlo simulation and extremum search by bisection algorithm.
As a result, the maximum flexibility power that a VB can provide with a
probabilistic guaranteed bound is obtained. The performance and validity of the
proposed method are demonstrated and discussed in three different case 
studies where a VB bids its aggregate power in the Spanish electricity
balancing markets (SEBM). 

% Keywords - if any
\end{abstract}
  
\section{Introduction} \label{introduction}
Electricity systems are undergoing a huge transformation due
to the need to switch from fossil energy sources to variable renewable energy
sources (RES).  At the same time, rising energy prices are putting electricity
markets as they are currently formulated to the test. The need to adapt the
demand curve to mitigate price peaks requires managing demand flexibility. To
achieve the desired transformation of the energy system, net metering and
self-consumption must be encouraged, and regulation is beginning to change
accordingly~\cite{poullikkas2013review,rehman2020penetration}.  At the same
time, demand is allowed to participate in new power markets and others
traditionally reserved for
generation~\cite{forouli2021assessment,gellings2016demand}.

In the residential sector, heating and cooling account for a large percentage
of energy demand \cite{fleiter2016mapping,sech2011analisis}. Most of the
appliances used for these purposes have a thermal inertia providing
flexibility, that is, their electricity demand can be shifted over time without
loss of utility. This is the case for thermostatically controlled loads (TCLs),
such as refrigerators, electric water heaters and heat pumps. The aggregation
of demand flexibility provided by many TCLs exhibits properties that
resemble those of batteries, hence this aggregation is referred to as a virtual
battery (VB). They have been extensively
studied~\cite{HHao1,SKhan,martin2021flexibility}.

VBs can be used to provide ancillary services \cite{zhao2016geometric} and to
participate in appropriate electricity markets \cite{song2019hierarchical}.
This requires accurate aggregate demand flexibility forecasting mechanisms.
Several approaches have been proposed to forecast demand flexibility.
In \cite{pinto2017multi} an evolutionary particle swarm optimisation combined
with support vector data description is used to find feasible trajectories for
residential loads.  
In \cite{lucas2019load} a combinatorial optimisation is proposed for
flexibility prediction using non-intrusive load monitoring in the residential
sector. Artificial neural networks are the main method employed in
\cite{macdougall2016applying,ponocko2018forecasting,merce2020load}.  
In \cite{hu2021flexibility}, a temporal convolution network was applied to
predict the flexibility of electric vehicle and domestic hot water systems.
Furthermore, the use of regression models based on machine learning is
discussed in \cite{ahmadiahangar2019residential}. 
Finally, a comprehensive review of flexibility prediction methodologies is
given in \cite{ahmadiahangar2020demand}.

Most of the approaches found in the literature do not take into account the
actual state of the loads when predicting the flexibility they can provide as a
whole, which can lead to inaccurate predictions. In addition, physical and
user-imposed constraints must also be taken into account to improve the
accuracy of predictions.

The main contribution of this work is the development of a novel method based
on Monte Carlo simulation and extremum search using the bisection method
to predict the flexibility that can be provided by a TCL aggregation
modeled as a VB. 
The key result for the development of the probabilistic method is formulated in
Theorem~\ref{th:prob} and allows estimating the probability of successfully
supplying a given constant power demand response with a guaranteed confidence
measure.

In addition, a method to control the VBs and the variability of their behavior is studied. Finally, the Spanish electricity balancing market (SEBM) is used
to test the potential of the proposed estimation method to facilitate demand
flexibility aggregators to make offers in this type of markets.

For predicting the flexibility, a function is introduced that measures the
probability that the VB will supply a given power over a period of time. This
function is estimated and optimized using Monte Carlo simulation and the
bisection method.
In addition, the minimum number of trials to be performed by Monte Carlo
simulation to obtain a probabilistic guarantee of the power that can be
supplied is obtained. Consequently, the proposed method estimates the maximum
power flexibility available in a given future time period with a probabilistic
guarantee measure.

The paper is organized as follows. Section~\ref{TCLandVB} explains the
materials an methods, specifically the TCLs and VBs models and their
management and control methodology. 
Besides, the electrical markets and the necessary
requirements for demand participation are also explained.
The main results are explained in detail in Section~\ref{flexpred}.
Here, the flexibility prediction is characterized by introducing a probability
function of the supply power of a VB. 
The probability of successfully supplying a given constant power during a
demand response event with a measure of confidence is obtained by applying the
result stated by Theorem~\ref{th:prob}. 
Based on this theoretical result, the MC\&ESB method is developed, which can be
easily applied by aggregators and energy service providers based on demand
response to make energy bids in the electricity market with high
confidence in their supply. 
In Section~\ref{casestud}, three case studies are discussed; the first one
estimates the likelihood function of a sample VB, the second one tests the
MC\&ESB method in three different scenarios in which VBs could participate in
SEBM, and the third one evaluates the response of the VB controller to a
requested power signal based on the SEBM results. 
Finally, Section~\ref{conclusion} concludes the paper.

%\section{TCL and VB modelling and control} \label{TCLandVB}
\section{Materials and methods} \label{TCLandVB}

The aggregate flexibility of a set of TCLs can be represented by a VB. The
management of the VB is enabled by the existence of a demand
flexibility control system that manages the operation of each of the TCLs in
the aggregation. In this work, the models and controller reported in
\cite{martin2021flexibility} are used due to the accuracy with which they are
able to satisfy the power requirements of the system operator by measuring the
current state of the appliances in real time. But, as explained below, some
improvements have been implemented in the control system that allow better
demand prediction and more flexibility in the aggregation response. The monitoring and communication system between TCLs and the controller is done using the architecture proposed in \cite{LAKSHMANAN2016705}.

\subsection{The TCL model}

A thermostatically controlled load (TCL) refers to a device or system
that, through a thermostat, automatically adjusts its operation based on temperature changes. The purpose of such a load is to regulate and maintain a
desired temperature within a specific range. TCLs are commonly used in heating,
ventilation, and air conditioning (HVAC) systems, as well as various appliances
and industrial processes, such as refrigerators and chemical reactors.

In the context of electrical systems, aggregations of TCLs can play a
significant role in optimizing energy consumption, demand response, and grid
stability. They can provide regulation services and mitigate power imbalances
resulting from fluctuating distributed renewable generation. 

A simple discrete-time model of a TCL is given by Equation \ref{eq:mod1} and Equation \ref{eq:mod2}. They
describe the time evolution of the internal temperature $\theta_{i}^{k}$,
which depends on its binary status $u_{i}^{k}$ (ON, OFF), the forecast ambient
temperature $\hat\theta_{a_{i}}^{k}$, and the disturbance $\omega_{i}^{k}$.
The parameters of the model are: the thermal resistance $R_{th_i}$, the thermal
capacity $C_{th_i}$, the nominal power $P_i$, and the performance coefficient
$\eta_i$. 

\begin{equation}\label{eq:mod1}
\theta_{i}^{k+1} =
g_{i} \cdot \theta_{i}^{k} +
(1 - g_{i}) \cdot (\hat\theta_{a_{i}}^{k} -
u_{i}^{k} \cdot \theta_{g_{i}}) + \omega_{i}^{k},
\end{equation}
where
\begin{equation}\label{eq:mod2}
g_{i} = e^{-1/{R_{th_{i}} \cdot C_{th_{i}}}}, \quad
\theta_{g_{i}} = R_{th_{i}} \cdot P_{i} \cdot \eta_{i}.
\end{equation}

The average power that each TCL is expected to demand $P_{0_{i}}^{k}$ is 
calculated using Equation \ref{eq:mod31}.

\begin{equation} \label{eq:mod31}
P_{0_{i}}^{k} = \frac{\hat\theta_{a_{i}}^{k}-\theta_{s_{i}}}{\eta_i \cdot R_{th_i}},
\end{equation}
where $\theta_{s_{i}}$ is the set point temperature of the TCL. The set point
temperature, $\theta_{s_{i}}$, and the band width $\Delta_i$
define the comfort band, or safety band, which is the range of temperatures
where the TCL must work and confers flexibility to the TCL, as described in Equation \ref{eq:mod55}.

\begin{equation} \label{eq:mod55}
\theta_{s_{i}} - \Delta_i \le \theta_{i}^{k} \le \theta_{s_{i}} + \Delta_i.
\end{equation}

The larger the safety band, the more flexibility the TCL will provide.

The model must also include several constraints that affect TCLs, such as
short-cycling prevention and availability. 
Short-cycling prevention avoids the TCL to change its status too frequently,
which could cause damage in the device components. The minimum status changing
frequency is specified by $\tau_{i}$. 
In addition, availability prevents the TCL from leaving the safety band and ensures that the
TCL does not operate when the ambient temperature is above the safety band in
heating appliances or below the safety band in cooling appliances.

\subsection{The VB model}

A VB models the aggregated demand flexibility of a set of TCLs at each time
instant $k$. The charging and discharging processes have different dynamics.
Consequently, two different variables are introduced in the model for the
evolution of energy and power magnitudes. Charging magnitudes are considered
positive, while discharging magnitudes are considered negative.

Let $m_{+}^{k}$ (resp. $m_{-}^{k}$) denote the maximum charging power 
(resp. discharging power) that can be provided by the VB at each instant. 
These variables measure whether a given flexible power demand can be supplied
by the TCL aggregation.
They are given by Equation \ref{rtrt} and Equation \ref{tyty}, respectively.

\begin{align} \label{rtrt}
m_{+}^{k+1} &= 
\sum_{i=1}^{N} (1-2\phi_{i}) \cdot (P_{i}-P_{0_{i}}^{k}) - P_{+}^{k}, \\
\label{tyty}
m_{-}^{k+1} &= \sum_{i=1}^{N} (1-2\phi_{i}) \cdot P_{0_{i}}^{k} - P_{-}^{k},
\end{align}
where $\phi_{i}$ is the device type, and $P_{+}^{k}$ and $P_{-}^{k}$ are the
the sum of power of the TCLs which are unavailable for charging or discharging,
respectively. 
See \cite{martin2021flexibility} for a more detailed explanation of the VB 
model and its applications.

\subsection{The demand flexibility control} \label{konami}

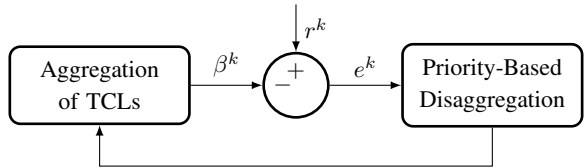
\begin{figure}
\centering
%\resizebox{\ifdim\width>\columnwidth\columnwidth\else\width\fi}{!}{
%\includegraphics[width=\columnwidth]{ImagenControlador}
%}
\begin{tikzpicture}
[
mybox/.style={
  rectangle,
  rounded corners,
  line width = 1pt,
  inner sep=6pt,
  text width=14ex,
  align=center,
  draw=black,
  fill=white
  },
mysum/.style={
  circle,
  line width = 1pt,
  inner sep=2pt,
  minimum size=6ex,
  align=center,
  draw=black,
  fill=white
  },
arrow/.style={
  ->,>=latex
  },
  scale=0.9, every node/.style={scale=0.9}
]
\node[mybox] (A) {Aggregation of TCLs};
\node[mysum, right= 6ex of A] (S) {};
\node[mybox, right= 6ex of S] (D) {Priority-Based Disaggregation};
\node[coordinate, above= 4ex of S] (R) {};
\node[coordinate, below= 4ex of S] (B) {};
\draw[arrow] (A) -- node[above]{$\beta^k$} (S);
\draw[arrow] (S) -- node[above]{$e^k$} (D);
\draw[arrow] (D) |- (B) -| (A);
\draw[arrow] (R) -- node[right]{$r^k$} (S);
\node[below = 1pt of S.north] {$+$};
\node[right = 1pt of S.west] {$-$};
\end{tikzpicture}
\caption{VB controller diagram \cite{martin2021flexibility}.}
\label{VBcontroller}
\end{figure}

The demand flexibility control developed for a VB operates in a cyclic 
process with two basic operations:
\textit{Aggregation of TCLs} and \textit{Priority-Based
Dissagregation}, as shown in Figure~\ref{VBcontroller}. 
The controller operates with a time step \textit{h}. As stated before, this control system is used in this paper because of its precision, checking the status of TCLs in real time.

At \textit{Aggregation of TCLs}, all the variables describing the current
situation of all TCLs are calculated, preventing their inner temperature from 
leaving the comfort band. Then, all the information of TCLs is
gathered. As a result, a deviation power signal $\beta^k$ is obtained, which is the the difference between the power consumed by the VB, its expected base consumption and the amount of power to be switched on or off in the next time instant because of TCLs constraints. Then, $\beta^k$ is compared with the system operator power signal $r^k$, which
sets the power requirement. This calculates the regulation power signal
$e^k$.

The \textit{Priority-Based Dissagregation} decides which TCLs must change their
status from on to off, or vice versa, in order to cope with the regulation
power signal $e^k$. The TCLs chosen first are the available ones that are far
in time from being switched on or off due to device constraints.

Additional improvements in the demand flexibility controller of \cite{martin2021flexibility} have been included
in this article. Firstly, the absolute error when tracking $r^k$ is reduced, as
the power difference between switching on or off the last TCL selected in
\textit{Priority-Based Dissagregation} is checked. In addition, the number of
$u_{i}^{k}$ changes are examined, and thus the activations of the short-cycle prevention
constraint are reduced. Secondly, the controller allows to configure periods of
time without demand management, which means that only \textit{Aggregation of
TCLs} step is executed. This is the case when no $r^k$ is required.

\subsection{Demand-side participation in SEBM} \label{flexmark}

Several European countries have opened their markets to consumers
participation, such us United Kingdom, Belgium, Germany or Sweden
\cite{forouli2021assessment, freire2022literature}. 
Here, we focus on Spanish electricity balance markets (SEBM), managed by Red
El\'ectrica de Espa\~na (REE).
In these markets, active power bids can be either upward or downward (from the
generation point of view). 
This is equivalent to a VB discharging, reducing consumption (negative
flexibility) or charging, demanding more power (positive flexibility),
respectively.

There are currently three SEBM: \textit{Secondary Regulation}, 
\textit{Tertiary Regulation} and 
\textit{Replacement Reserves} \cite{BOEref}. 

\textit{Secondary Regulation (SR)} 
is an optional ancillary service whose purpose is to
maintain the balance between generation and demand, correcting the
unintentional deviations.
Its temporal working horizon ranges from 30
seconds to 15 minutes. 

\textit{Tertiary Regulation (TR)}
is an optional ancillary service that, if subscribed to, is accompanied by the
obligation to bid and is managed and compensated by market mechanisms. Its
objective is to resolve the deviations between generation and consumption and
the restitution of the secondary control reserve which has been used. 

\textit{Replacement Reserves (RR)} 
is an optional service managed and remunerated by market mechanisms. The
objective is to resolve the deviations between generation and demand which
could appear in the period between the end of one intraday market and the
beginning of the next intraday market horizon.

All of them focus on maintaining the balance between generation and demand, and operate in fifteen-minute periods. 
The bids submitted to the markets consist mainly of the active power offered, the price, and some possible complex constraints, such as indivisibility. 
Once the bids are submitted, a matching algorithm is executed. 
The SEBM comply with the European Commission Regulation \cite{EUref} and
ENTSO-E nomenclature \cite{ENTSOEweb}.  
The main characteristics of the three SEBM are given in Table~\ref{SbaMa}.

\begin{table*}[!t] 
\caption{SEBM characteristics \cite{BOEref,marsboom2018proposal}.}
\label{SbaMa}
\centering
\resizebox{\ifdim\width>\textwidth\textwidth\else\width\fi}{!}{
\begin{tabular}{ccccccc}
\toprule
Market name & 
\textit{Secondary Regulation (SR)} &
\textit{Tertiary Regulation (TR)} & 
\textit{Replacement Reserves (RR)} \\
\midrule
ENTSO-E nomenclature & 
\begin{tabular}{c}
Automatic Frequency \\
Restoration Reserves (aFRR) 
\end{tabular} &
\begin{tabular}{c}
Manual Frequency \\
Restoration Reserves (mFRR)
\end{tabular} &
Replacement Reserves (RR) \\
European platform & PICASSO & MARI & TERRE \\
Mode of activation & Automatic & Manual & Manual \\
Maximum ramping period & Real-time & 15 min & 30 min \\
\begin{tabular}{c}
Minimum time in advance \\
for submission of bids
\end{tabular} & 
16:00 D-1 & 25 min & 55 min \\
\bottomrule
\end{tabular}
}
\end{table*}

All the requirements that demand aggregators need to participate in SEBM are
gathered in \cite{REEguia}. 
According to this, the minimum power bid must be 1~MW.

The method developed in this paper encourages demand-side participation because
it provides demand aggregators with a rigourous tool to calculate how
much power they could offer in electricity markets.

%\section{Demand flexibility prediction} \label{flexpred}
\section{Results} \label{flexpred}

The control strategy explained in the previous section allows a one-step
ahead prediction of the VB parameters, as $m_{+}^{k}$ and $m_{-}^{k}$ are
obtained before $r^{k}$. However, this forecasting horizon is not enough for
sophisticated demand flexibility management, including participation in
electricity balancing markets, where the demand flexibility must be
accurately predicted before the actual provision of the service and for 
longer prediction horizons. The novel MC\&ESB method presented in this paper solves this issue, as the predictions of the method can be calculated several hours in advance.

\subsection{The VB power supply probability function}

Predicting the flexibility of VBs requires studying how the temperatures of
TCLs are expected to change over time. However, this is not an easy task. The
evolution of TCL temperatures is influenced by errors in ambient temperature
forecasting, model inaccuracy and other disturbances related to the operation
of the devices (e.g. when a refrigerator is opened).
Representing these inaccuracies and perturbations requires the use of random
models. 
The TCLs used in this work encapsulate all of this uncertainty in the
perturbation parameter $\omega_{i}^{k}$. Usually, $\omega_{i}^{k}$ is
considered to be distributed as a Gaussian random variable of zero expectation
and constant variance $\sigma^2$, see \cite{mathieu2013energy}.

The randomness in the temperature evolution of TCLs is a consequence of random
disturbances and uncertainty in the knowledge of the initial conditions.
Consequently, the ability of a VB to respond to a given flexibility demand,
i.e. to satisfy a given power deviation from the reference consumption during a
given period of time $t$, is a random variable characterized by a given
probability distribution.

The VB is composed of a set of TCLs. The number of elements in this set, $N_{TCL}$, is considered large
enough to achieve generality and reduce the influence of $\omega_{i}^{k}$. 

Let $x$ be a real-valued variable representing a candidate deviation power that
the VB could provide during a given time period $t$. 
In order to characterize the capability of VB to provide a power
deviation during certain time period, a power supply probability function 
$\Phi(x)$ is defined.

Let $S$ be a binary random variable representing whether a constant power 
can be successfully provided by the VB for a certain period of time. 
The binary random variable takes the value $S=1$ if the 
VB successfully provides the demanded power. Otherwise, $S=0$.

Let $\Phi(x)$ be the probability that the VB can successfully supply a
constant power of value $x$ during a time horizon $t$. The function 
$\Phi(x)$ is a likelihood function and is given by the conditional probability
\begin{align}  \label{rtrt5}
\Phi(x) = \mathbb P [ S = 1 \mid x ]
\end{align}

In spite of the random nature of the VBs system behavior, we can assume that
the probability function $\Phi(x)$ is monotonic when the number of TCLs 
in the VB is large. In other words, the larger the absolute value of 
the demanded power to a VB for a given duration, the lower the 
probability of actually supplying it.

\subsection{The MC\&ESB method for estimating a VB power supply probability
function} \label{tiruri}

The monotonicity property of the power supply probability function $\Phi(x)$ can be used to estimate the maximum power (positive or negative) that a given VB could supply to an aggregator or other market actor with a probabilistic guarantee measure. This estimate is a prediction of the demand flexibility that can be provided by a TCL aggregate that is modeled as a VB.

To find the maximum flexible demand power that a VB can supply, a new method,
named MC\&ESB, is designed in this paper. It is based on the combination of
Monte Carlo simulation techniques and extremum search using the bisection
algorithm.

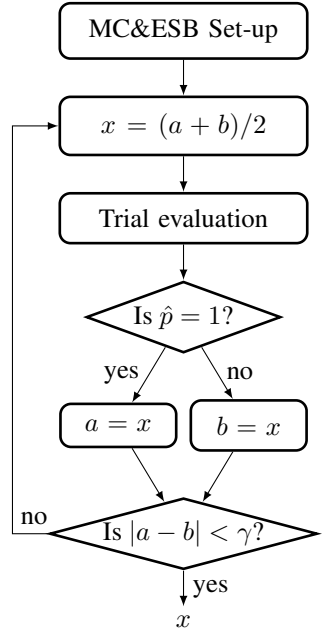
\begin{figure}
\centering
%\resizebox{\ifdim\width>\columnwidth\columnwidth\else\width\fi}{!}{
%\includegraphics[width=0.75\textwidth]{ImagenControlador}
%\includegraphics[width=\columnwidth]{ImagenMCESB}
%}
\begin{tikzpicture}
[
mybox/.style={
  rectangle,
  rounded corners,
  line width = 1pt,
  inner sep=6pt,
  text width=18ex,
  align=center,
  draw=black,
  fill=white
  },
mysum/.style={
  circle,
  line width = 1pt,
  inner sep=2pt,
  minimum size=6ex,
  align=center,
  draw=black,
  fill=white
  },
mydec/.style={
  diamond,
  aspect=4,
  line width = 1pt,
  inner sep=0pt,
  minimum size=6ex,
  align=center,
  draw=black,
  fill=white
  },
arrow/.style={
  ->,>=latex
  }
]
\node[mybox] (A) {MC\&ESB Set-up};
\node[mybox,below= 3ex of A] (B) {$x=(a+b)/2$};
\node[mybox,below= 3ex of B] (C) {Trial evaluation};
\node[draw,mydec,below= 3ex of C] (D) {Is $\hat p = 1$?};
\node[coordinate,below= 6ex of D] (E) {};
\node[mybox,text width = 7ex,left= 0.5ex of E] (F) {$a = x$};
\node[mybox,text width = 7ex, right= 0.5ex of E] (G) {$b = x$};
\node[draw,mydec,below= 6ex of E] (H) {Is $|a-b| < \gamma$?};
\node[below= 3ex of H] (I) {$x$};
\draw[arrow] (A) -- (B);
\draw[arrow] (B) -- (C);
\draw[arrow] (C) -- (D);
\draw[arrow] (D) -- node[left] {yes} (F);
\draw[arrow] (D) -- node[right] {no} (G);
\draw[arrow] (F) -- (H);
\draw[arrow] (G) -- (H);
\draw[arrow] (H) -- node[right]{yes} (I);
\draw[arrow] (H.west) -| ++(-3ex,0) node[above,pos=.2]{no} |- (B);
\end{tikzpicture}

\caption{MC\&ESB method diagram \cite{martin2021flexibility}.}
\label{zelda}
\end{figure}

Monte Carlo simulation is a probabilistic method useful for estimating a value
under uncertainty, especially for complex systems
\cite{papadopoulos2001uncertainty}. It consists of evaluating a function a
large number of times, so it finally converges to the most probable solution
(the mean value) despite of the existing uncertainty \cite{zhang2021modern}.
The accuracy of the method increases as the number of evaluations of the
function increases \cite{landau_binder_2014}.

We consider that a VB can provide the requested flexibility when the
probability of supplying the aggregate power target for a given time period $t$
is equal to 1 with a certain accuracy $\epsilon$ and confidence $\delta$. A VB
supplies the power target whenever $|r^k|$ is less than or equal to
$|m_{+}^{k}|$ for positive flexibility or $|m_{-}^{k}|$ for negative
flexibility.

Let $p$ be the probability that the virtual battery provides a power target of
value $x$, then $\Phi(x)=\mathbb P[S=1|x]=p$. This probability $p$ is constant
but unknown, but can be estimated by performing a sequence of experiments. The
confidence measure of the estimate can also be obtained from experiments using
Bayesian inference.

Before any trial is performed, our belief about the probability $p$ is modeled
by a uniform distribution in the unit interval $p \sim U[0,1]$.
Each trial $j$ determines whether or not the virtual battery can supply  power
$x$ under random initial conditions for each individual TCL in the VB. 
Thus, the outcome of each trial is a binary random variable $\Xi_j$ that takes
the value 1 if the power deviation $x$ is supplied during the demand response
event duration $t$, and 0 otherwise. 
The experiments are statistically independent of each other, i.e. they are
Bernouilli's trials.
Given a sequence of $N$ Bernouilli's trials $\{\Xi_j \mid j=1,\ldots,N\}$, the
random variable $\Xi = \sum_{j=1}^{N} \Xi_j$ is distributed according to a
binomial distribution $B(N,p)$ with conditional probability function
\begin{align} \label{rtert}
\mathbb P(\Xi=n \mid p) = {N \binom{n}{m} n} p^n (1-p)^{N-n}, \ p\in [0,1]
\end{align}
The conditional probability $\mathbb P(\Xi=n \mid p)$ is called the likelihood
function and is a function of $p$, 
\emph{i.e.} $L(p) = \mathbb P(\Xi=n \mid p)$.

The probability of $p$ conditioned on the outcome of the experimentation process
is called \emph{a posteriori} probability and can be obtained by applying
Bayes' Theorem
\begin{align}
f(p \mid \Xi=n) 
&= \frac{\mathbb P(\Xi=n \mid p)f(p)}{\mathbb P(\Xi=n)} 
\nonumber \\
&= \frac{{N \binom{n}{m} n} p^n (1-p)^{N-n}}{\mathbb P(\Xi=n)} 
\nonumber \\
&= \frac{1}{c(n,N)} p^n (1-p)^{N-n}
\end{align}
where $c(n,N)=\int_0^1 p^n(1-p)^{N-n} dp$. 

The estimate of the probability $p$ can be obtained from the 
\emph{a posteriori} probability density function as
\begin{align}
\hat p = \arg\max_{p\in[0,1]} f(p \mid \Xi=n) 
\end{align}
which is
\begin{align} \label{ecuecu}
\hat p = \frac{n}{N}
\end{align}
and a confidence measure of this estimate 
also is obtained using the \emph{a posteriori} probability
density funcion of $p$ 
\begin{align} \label{eq:confbeta}
\mathbb P
\left\{ \frac{n}{N}-\epsilon_1 \leq p \leq  \frac{n}{N}+\epsilon_2 \right\} 
= \quad\quad\quad
\nonumber \\
\frac{1}{c(n,N)} 
\int_{\frac{n}{N}-\epsilon_1}^{\frac{n}{N}+\epsilon_2} p^n (1-p)^{N-n} dp
\end{align}
where $c(n,N)=\int_0^1 p^n (1-p)^{N-n} dp$.

Using the above expression, we are interested in obtaining the number of
trials to estimate the interval of power range $[x_{\min},x_{\max}]$
that can be supplied during the demand response event of duration $t$ with a 
given confidence.

\begin{theorem}\label{th:prob}
Let $\epsilon$ and $\delta$ be scalars in the open unit interval $(0,1)$
then 
\begin{align}
\mathbb P(p \geq 1-\epsilon \mid \Xi = N) \geq 1-\delta
\end{align}
whenever
\begin{align}
N \geq \frac{\ln (1/\delta)}{\ln (1/(1-\epsilon))}-1 \label{clave}
\end{align}
\end{theorem}

\paragraph*{Proof.} 
If $\Xi=N$, then the \emph{a posteriori} density function of $p$ is given by
$f(p \mid \Xi=N) = (N+1)p^N$. Then,
\begin{align*}
\mathbb P (p \geq 1-\epsilon \mid \Xi=N) 
&= \int_{1-\epsilon}^1 f(p \mid \Xi=N) dp \\
&= \int_{1-\epsilon}^1 p^N dp \\
&= (N+1) \left[ \frac {p^{N+1}}{N+1} \right]_{1-\epsilon}^{1} \\
&= 1-(1-\epsilon)^{N+1}
\end{align*}
Since  $N \geq \frac{\ln (1/\delta)}{\ln (1/(1-\epsilon))}-1$ is equivalent
to $1-(1-\epsilon)^{N+1} \geq 1-\delta$, the result is proved.
\hspace*{\fill}$\Box$

The above theorem allows us to estimate the probability of successfully
supplying a given constant power during the demand response event with a
measure of confidence.  
Let $x_{\max}$ be the maximum value of $x$ such that $\Xi=N$, then 
$p \geq 1-\epsilon$ with probability greater that $1-\delta$ if no trial fails
for a number of trials satisfying
$N \geq \frac{\ln (1/\delta)}{\ln (1/(1-\epsilon))}-1$.
In a similar way,
let $x_{\min}$ be the minimum value of $X$ such that $\Xi=N$, then
$p \geq 1-\epsilon$ with probability greater that $1-\delta$ if no trial
fails for a number of trial satisfying
$N \geq \frac{\ln (1/\delta)}{\ln (1/(1-\epsilon))}-1$.

In the Monte Carlo simulation stage of the MC\&ESB method, the time evolution of the VB is simulated for each candidate value of power $x$. An estimate of the probability $p$ is given by Equation \ref{ecuecu}, where $N$ is obtained
using Equation~\ref{clave}.

A schematic graphical description of the method is shown in Figure
\ref{zelda}. The stage of the extremum search using the bisection algorithm in the MC\&ESB
method consists of searching for the maximum value of the power $x$ in absolute
value that the VB can supply with probability 1. The method starts by defining
a search interval, which must be between $[0,+\infty]$ when searching for
flexibility in charging and between $[-\infty,0]$ when searching for
flexibility in discharging. Next, the bisection algorithm computes a first
value of $x$ and the estimate of the probability $p$ is computed by 
Equation~\ref{ecuecu}.
The bisection interval bounds $a$ and $b$ are updated at each iteration. The
bisection algorithm terminates when $|a - b| < \gamma$, with $\gamma$ being a
given tolerance.
The solution found $x$ is the maximum flexible demand power that the VB can
supply with probability greater than $1-\epsilon$ and confidence $1-\delta$.

In this paper, the TCL and VB modelling and control presented in Section~\ref{TCLandVB} has been used as evaluation function for MC\&ESB, but any other could be used as long as the effect of TCLs disturbances are considered.

%\section{Case Studies} \label{casestud}
\section{Discussion} \label{casestud}

This section discusses three case studies. The first case estimates the power
supply probability function describing the behavior of a VB composed of a set
of TCLs. The second case tests the performance of the MC\&ESB method for
participating in SEBM in different scenarios. The third case shows how the
control of a VB manages the aggregated set of TCLs to supply the power needed
to bid in SEBM.

\subsection{Case study 1: Estimation of VB supply probability function}

In this case study, the power supply probability function $\Phi$
is experimentally estimated along with confidence intervals for a 
certain VB composed of 3000 TCLs.

The TCLs participating in the VB are 1000 refrigerators, 1000 electric water
heaters and 1000 reversible heat pumps, which are considered as heating pumps
in winter and cooling pumps in summer. Their characteristics are shown in Table
\ref{typicalpara}. 
The location of the VB is Madrid, in Spain, and the season is summer.
Thus, The forecast ambient temperature of every device $i$,  
$\hat\theta_{a_{i}}^{k}$, is considered constant and equal to 
$24$\textdegree{}C
for refrigerators and electric water heaters, whereas for
reversible heat pumps is the temperature of August, 10th at 15:00 UTC of
the typical meteorological year, obtained with PVGIS \cite{huld2012new}. 

The VB aims to modulate its consumption with a constant $r^k$,
and the demand response event $t$ is equal to 15 minutes, which is the time
period in which SEBM operates. 

The initial status of each devices $u_{i}^{0}$ has been
randomized, the variance $\sigma^2$ is set to $0.05$,
the minimum status changing frequency $\tau_{i}$ is 1, and the control
time step $h$ is 1 minute (1/60 hours). 

\begin{figure}
  \centering
  \resizebox{\ifdim\width>\columnwidth\columnwidth\else\width\fi}{!}{
    \includegraphics[width=\textwidth]{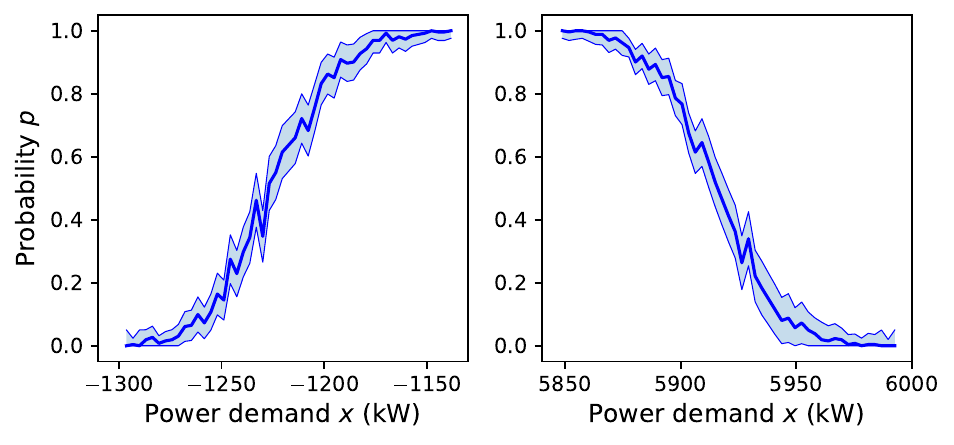}
  }
  \caption{Estimated demand flexibility probability function $\Phi(x)$ and its confidence bounds for the VB of case study 1}
  \label{fig:probcurv}
\end{figure}

The accuracy and confidence parameters $\epsilon$ and $\delta$ have been set to
0.02 and 0.005, respectively. Then, applying Theorem~\ref{th:prob}, the
number of trials is $N=262$. 
The MC\&ESB method is applied to obtain $x_{\min}$ and $x_{\max}$, 
where $x_{\min}$ (resp. $x_{\max}$) is the minimum (resp. maximum) value of 
power that the VB can always supply for $N=262$ experiments.  
Therefore, the probability of supplying power $x$ for any 
$x\in[x_{\min},x_{\max}]$ is greater that $0.98$ with confidence 
at least of $0.995$.
For this VB, $x_{\min}=-1138.3$~{kW} and $x_{\max}=5841.8$~{kW}.
By analogy, the MC\&ESB method is also applied to obtain 
$x'_{\min}$ and $x'_{\max}$, where $x'_{\min}$ (resp. $x'_{\max}$) is the limit
value of power such that if $x\leq x'_{\min}$ (resp. $x\geq x'_{\max}$),
the VB can never supply the power $x$ for $N=262$ experiments. 
Therefore, the probability of not supplying power $x$ for any 
$x\not\in[x'_{\min},x'_{\max}]$ is greater that $0.98$ with confidence 
at least of $0.995$.
For this VB, $x'_{\min}=-1296.2$~{kW} and $x'_{\max}=5992.6$~{kW}.
For $x\in[x'_{\min},x_{\min}]$ (resp. $x\in[x_{\max},x'_{\max}]$) the 
probability function $\Phi$ and their bounds for a confidence of at
least $0.995$ are depicted in Figure~\ref{fig:probcurv}. The curves have
been obtained for $50$ equally spaced supply powers $x$. 
At each power $x$, the probability
is estimated as $\hat p = n/N$, where $n$ is the number of experiments 
where the power $x$ is successfully supplied and $N=262$. The lower and
upper bounds $\hat p-\epsilon_1$ and $\hat p+\epsilon_2$ have been obtained
using Equation~\ref{eq:confbeta} such that
$\mathbb P[\hat p - \epsilon_1 \leq p \leq \hat p + \epsilon_2]=1-\delta=0.995$.
The bounds $\epsilon_1$ and $\epsilon_2$ have been selected to be equal,
whenever possible.
However, this is not possible when $\hat p$ approaches
$0$ or $1$. Consequently, non symmetric bounds are considered for these cases.
The estimated demand probability function $\Phi$ and its confidence bounds 
can be used to characterize the capability of a VB. For example, for the 
VB of this study, we can state that it can supply a power 
$x \in [-1138.3,5841.8]$ during $15$ minutes 
with probability greater than $0.98$ and confidence $0.995$. 
Moreover, we can also state from the data represented in 
Figure~\ref{fig:probcurv} that the VB can supply a power $x = 5900$~kW during 
$15$ minutes with probability $p \in [0.7021,0.8322]$ and confidence $0.995$.
Thus, the knowledge of the power supply probability function $\Phi$ for a given
VB is crucial to decide about participating in electrical markets.

\begin{table*}[!t]
\caption{Range of values for the parameters of residential~TCLs \cite{mathieu2012modeling} (pump set-points from \cite{BOEref2}).}
\label{typicalpara}
\centering
\resizebox{\ifdim\width>\textwidth\textwidth\else\width\fi}{!}{
\begin{tabular}{ccccccc}
\toprule
TCL Type & $R_{th}$ (\textdegree C/kW) & $C_{th}$ (kWh/\textdegree C) & $P$ (kW) & $\eta$ & $\theta_s$ (\textdegree C) & $\Delta$ (\textdegree C) \\
\midrule
Heating pumps & $2$ & $2.0$ & $-5.6$ & $3.5$ & $22.0$ & $0.5$ \\
Cooling pumps & $2$ & $2.0$ & $5.6$ & $2.5$ & $24.0$ & $0.5$  \\
Electric water heater & $120$ & $0.4$ & $-4.5$ & $1.0$ & $48.5$ & $3.0$ \\
Refrigerator & $90$ & $0.6$ & $0.3$ & $2.0$ & $2.5$ & $1.5$ \\
\bottomrule
\end{tabular}
}
\end{table*}

\subsection{Case study 2: Demand flexibility prediction}

The second case study aims to forecast the maximum flexibility power that the 
same VB of case~1 can supply at different times of the year to participate in
the different SEBM markets using the MC\&ESB method. Three scenarios
have been considered.  
In all of them, the simulation is divided into two periods.

The duration of the first period is the sum of the time taken by the algorithm
to calculate the prediction (e.g. 5 min) and the minimum time required for the
submission of bids in the corresponding SEBM. The duration of the second period
corresponds to the demand management horizon, $t$ equal to $15$ minutes.
The first period simulates the TCL evolution without any requested $r^k$ power signal, while in the second period demand management is activated.

The MC\&ESB method is configured to search for the maximum flexibility power 
with probability greater than $0.98$ and confidence $0.995$, as explained in 
Section \ref{tiruri}.
The remaining variables, parameters and VB characteristics, if not mentioned
below, are the same as in case study 1.

In Scenario 1, the aggregator wants the VB to participate in 
\textit{Secondary Regulation} from 01:00 to 01:15 UTC on 7th February. 
At those hours, only refrigerators and electric water heaters operate. 
The requested flexibility power is positive because low power consumption 
is expected at night. 
The ambient temperature $\hat\theta_{a_{i}}^{k}$ is considered constant and
equal to $22$ \textdegree C. the bounds of the bisection algorithm
$a$ and $b$ are $0$ and $5000$, respectively.

Scenario 2 takes place in a summer afternoon. The VB is expected to participate
in \textit{Tertiary Regulation} from 16:00 to 16:15 UTC on 19th July. In this
case, all the reversible heat pumps operate as cooling pumps. The requested
flexibility power is negative to achieve peak saving, as the power consumption
on the grid is expected to be high. The ambient temperature
$\hat\theta_{a_{i}}^{k}$ is considered constant and equal to 24 \textdegree C
for refrigerators and electric water heaters, while that of the cooling pumps
is obtained from the typical PVGIS weather year. The bounds of the bisection
algorithm $a$ and $b$ are $0$ and $-3500$, respectively.

Finally, Scenario 3 considers the morning of a winter day. The VB is intended
to participate in \textit{Replacement Reserves} from 10:00 to 10:15 UTC on 5th
January. The requested flexibility power is negative, as in Scenario 2. The
ambient temperature $\hat\theta_{a_{i}}^{k}$ is considered constant and equal
to 22 \textdegree C for refrigerators and electric water heaters, while for
heating pumps it is obtained from the typical PVGIS weather year. The bounds
$a$ and $b$ are $0$ and $-3500$, respectively.

\begin{table}[!t] 
\caption{Flexibility power available at each scenario (kW).}
\label{rrrrt}
\centering
\begin{tabular}{ccc}
\toprule
Scenario 1 & Scenario 2 & Scenario 3 \\
\midrule
3608.0 & -1482.9 & -1250.1 \\
\bottomrule
\end{tabular}
\end{table}

The resulting flexibility power at each scenario is showed in Table~\ref{rrrrt}.
The highest amount of power flexibility is obtained in Scenario~1, Followed by
Scenario~2 and Scenario~3. Usually, the VB can provide more positive than
negative power flexibility because TCLs take longer to increase or decrease
their temperature when this change is not caused by an electromechanical
element \cite{martin2021flexibility}. All calculated powers are above 1~MW in
absolute value, which allows the aggregator to make bids in the SEBMs. 
These results demonstrate that the MC\&ESB method is useful to calculate the
availability of flexible power to participate in SEBMs and other similar
markets in a short period of time, depending on the computational capabilities.

\subsection{Case study 3: Demand flexibility control}

\begin{figure} \
  \centering
  \resizebox{\ifdim\width>\columnwidth\columnwidth\else\width\fi}{!}{
    \includegraphics[width=\textwidth]{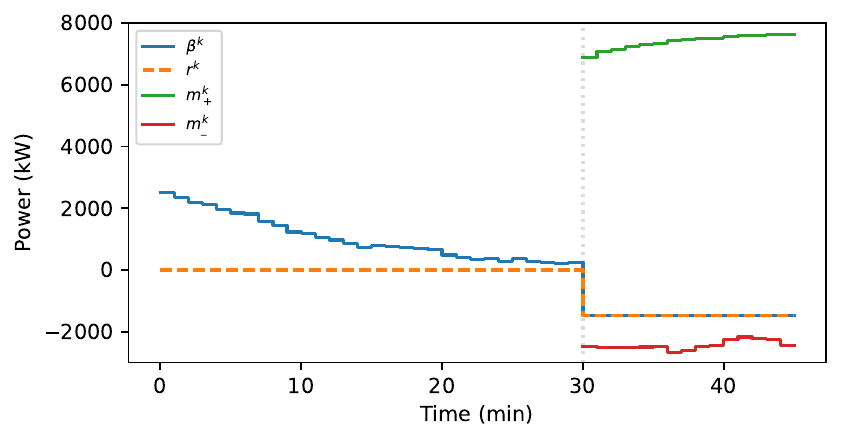}
  }
  \caption{VB deviation power. At minute 30 the demand flexibility control starts to actuate.}
  \label{re1}
\end{figure}

\begin{figure}
  \centering
  \resizebox{\ifdim\width>\columnwidth\columnwidth\else\width\fi}{!}{
    \includegraphics[width=\textwidth]{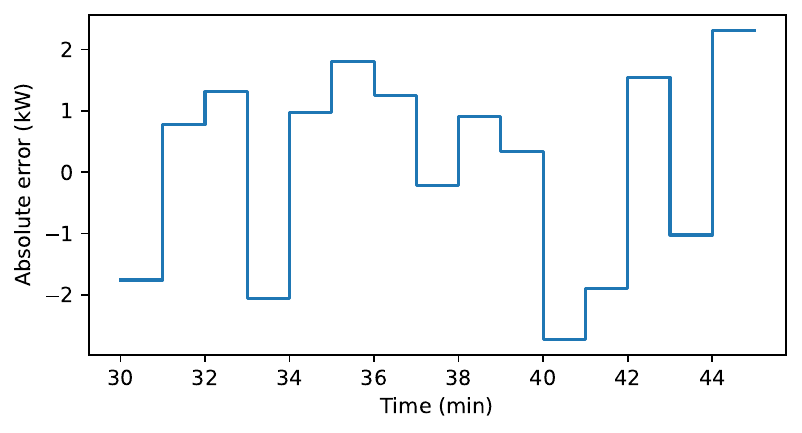}
  }
  \caption{Absolute error of the VB controller.}
  \label{re2}
\end{figure}

This case study shows how flexibility is managed once the aggregator's offer
has been accepted in the market and it is time to supply the energy. 
The flexibility power obtained in Scenario 2
of the previous case study ($-1482.9$ kW) must now be supplied to the grid. For
this purpose, the demand flexibility control system discussed in Section
\ref{konami} is activated. The VB has been simulated with the same conditions
as in Scenario 2 during 45 minutes. During the first 30 minutes the TCLs are
not managed by the controller, while in the last 15 minutes the controller
comes into operation to supply the requested power. The operation is shown in
Figures~\ref{re1} and \ref{re2}.

The VB tracks the $r^k$ power signal as soon as the controller is activated,
i.e. from the 30th minute onwards. Before that, the TCLs consume the electrical
power they need to satisfy the users' comfort needs without any restriction.
This comfort band is also maintained at all times during the 15 minutes of
controller operation, but the coordination in the state of the TCLs makes it
possible to supply the requested power flexibility.
The VB consumption during the first 30 minutes was higher than the expected
baseline consumption. For this reason, $\beta^k$ is greater than 0 during this
period. Deviations from the consumption baseline could occur  in real time, so
it is important to know the current state of the devices when managing
flexibility.

Two aspects are worth noting. The first one is that $\beta^k$ is always between
$m_{+}^{k}$ and $m_{-}^{k}$ during controller operation, which means that the
VB responds appropriately  to the control signal $r^k$. The second one is
illustrated in Figure \ref{re2}, and refers to the fact that the absolute error
of the controller is never larger than half of the maximum $P_i$ of the TCLs
participating in the VB, provided that $r^k$ is always feasible.

\section{Conclusion} \label{conclusion}

Demand-side participation in the electricity system is key to reducing the use
of fossil fuels and increasing the penetration of renewables in the system.  In
this paper, a new method for predicting demand flexibility, called MC\&ESB, has
been developed. The predictability provided by MC\&ESB allows an aggregator to
participate in electricity markets where demand flexibility can be traded. The method obtains accurate predictions as it can be used with a controller which takes into account the actual state of the loads, as in this paper. Nevertheless, any other control system could be utilized, provided that load disturbances are taken into consideration. The
results of demand management with the newly developed approach have been
illustrated in three case studies in the Spanish ancillary services electricity
market. As a result, the flexibility power above 1 MW has been predicted and
the control of the VBs has been simulated.

The VBs used in this work are composed of TCLs, but other sources of residential flexibility, such as
electric vehicles and household batteries, could be considered. In addition,
the VBs are not limited to residential loads. The MC\&ESB method could be used
in the industrial sector. Future work will focus on improving the TCL models
and testing the methodology in a real environment.

%\section{Nomenclature}

\nomenclature{$i$}{A thermostatically controlled load}
\nomenclature{$k$}{A time instant [h]}
\nomenclature{$u_{i}^{k}$}{Status of $i$ at $k$ (0 off, 1 on)}
\nomenclature{$\theta_{i}^{k}$}{Internal temperature of $i$ at $k$  [\textdegree C]}
\nomenclature{$\hat\theta_{a_{i}}^{k}$}{Forecast ambient temperature of $i$ at $k$ [\textdegree C]}
\nomenclature{$R_{th_i}$}{Thermal resistance of $i$ [\textdegree C/kW]}
\nomenclature{$C_{th_i}$}{Thermal capacitance of $i$ [kWh/\textdegree C]}
\nomenclature{$P_i$}{Nominal power of $i$ ($+$ cooling, $-$ heating) [kW]}
\nomenclature{$\eta_i$}{Coefficient of performance of $i$}
\nomenclature{$\omega_{i}^{k}$}{Disturbance of $i$ at $k$ [\textdegree C]}
\nomenclature{$P_{0_{i}}^{k}$}{Average expected demanded power of $i$ at $k$ [kW]}
\nomenclature{$\theta_{s_{i}}$}{Set point temperature [\textdegree C]}
\nomenclature{$\Delta_i$}{Temperature dead band of $i$ [\textdegree C]}
\nomenclature{$\tau_{i}$}{Minimum status changing frequency of $i$ [h]}
\nomenclature{$\phi_{i}$}{Device type of $i$ (0 cooling, 1 heating)}
\nomenclature{$N_{TCL}$}{Number of aggregated thermostatically controlled load}
\nomenclature{$m_{+}^{k}$}{Maximum available charging power at $k$ [kW]}
\nomenclature{$m_{-}^{k}$}{Maximum available discharging power at $k$ [kW]}
\nomenclature{$P_{+}^{k}$}{Power unavailable for charging at $k$ [kW]}
\nomenclature{$P_{-}^{k}$}{Power unavailable for discharging at $k$ [kW]}
\nomenclature{$h$}{Control time step [h]}
\nomenclature{$\beta^k$}{Deviation power signal [kW]}
\nomenclature{$r^k$}{System operator power signal [kW]}
\nomenclature{$e^k$}{Regulation power signal [kW]}
\nomenclature{$\sigma^2$}{Variance of Gaussian disturbance [\textdegree C]}
\nomenclature{$t$}{Demand management event duration [h]}
\nomenclature{$x$}{A value of supply power [kW]}
\nomenclature{$p,\hat p$}{Supply power probability value and its estimate}
\nomenclature{$\Phi,\hat\Phi$}{Supply power probability function and its estimate}
\nomenclature{$L(\cdot)$}{Likelihood function}
\nomenclature{$S$}{Capability to supply certain power (0 no, 1 yes)}
\nomenclature{$j$}{A Bernouilli trial}
\nomenclature{$\Xi_j$}{Result of $j$ (0 no, 1 yes)}
\nomenclature{$\Xi$}{Sum of $\Xi_j$}
\nomenclature{$N$}{Number of Bernouilli trials}
\nomenclature{$n$}{Number of successful Bernouilli trials}
\nomenclature{$\epsilon,\epsilon_1,\epsilon_2$}{Probability bounds}
\nomenclature{$\delta$}{Probability confidence}
\nomenclature{$x_{\max}$}{Maximum power with guaranteed probability [kW]}
\nomenclature{$x_{\min}$}{Minimum power with guaranteed probability [kW]}
\nomenclature{$x'_{\max}$}{Maximum power with nonzero probability [kW]}
\nomenclature{$x'_{\min}$}{Minimum power with nonzero probability [kW]}
\nomenclature{$a,b$}{Bounds of the bisection search algorithm [kW]}
\nomenclature{$\gamma$}{Tolerance of the bisection search algorithm [kW]}
\printnomenclature

\section*{Data availability}

Data will be made available on request.

\section*{Conflicts of interest}

The authors declare that they have no conflicts of interest.

\section*{Funding statement}

This research has been partially supported by the European Union ERDF (European
Regional Development Fund) and the Junta de Castilla y Le\'on
through ICE (Instituto para la Competitividad Empresarial) to improve
innovation, technological development, and research, dossier no.
CCTT1/17/VA/0005.

% Bibliography style - if using a .bib file
	\bibliographystyle{hindawi_bib_style}
	\bibliography{cas-refs} % without .bib extension

\end{document}